\documentclass[twocolumn,natbib,runningheads]{svjour3}
\usepackage{graphicx}
\journalname{Astrophysics and Space Science}
\smartqed

\begin{document}
\author{\v{S}. Parimucha \and T. Pribulla \and M. Va\v{n}ko \and P. Dubovsk\'y \and L. Hamb\'alek}

\title{Photometric Analysis of Recently Discovered Eclipsing Binary GSC~00008-00901}

\titlerunning{Photometric Analysis of GSC 00008-00901}

\authorrunning{Parimucha et al.} 

\institute{\v{S}. Parimucha \at
              Institute of Physics, Faculty of Natural Sciences, \v{S}af\'{a}rik University Ko\v{s}ice, Slovakia \\
              \email{parimuch@ta3.sk}
           \and
              T. Pribulla, M. Va\v{n}ko, L. Hamb\'alek \at
              Astronomical Institute of Slovak Academy of Sciences,
              Tatransk\'{a} Lomnica, Slovakia\\
              \email{(pribulla, lhambalek)@ta3.sk}
           \and
           M. Va\v{n}ko  \at {Astrophysikalisches Institut, Universit\"{a}t Jena, 
                Schillerg\"{a}sschen 2-307745 Jena,  Germany\\
               \email{vanko@astro.uni-jena.de}
            \and
           P. Dubovsk\'y \at
              Kolonica Observatory, Slovakia\\
              \email{var@kozmos.sk}
}}

\date{Received: date / Accepted: date}

\maketitle

\begin{abstract}
Photometric analysis of $BVR_C$ light curves of newly discovered eclipsing binary GSC~0008-00901 is presented. The orbital period is improved to 0.28948(11) days. Photometric parameters are determined, as well. The analysis yielded to conclusion that system is an over-contact binary of W UMa type with components not in thermal contact. The light curves from 2005 show the presence of a spot on the surface of one of the components, while light curves from 2006 are not affected by maculation.


\end{abstract}

\section{Introduction}

The variability of GCS 00008-00901 ($\alpha_{2000}$= 00$^{\mathrm h}$ 13$^{\mathrm m}$ 22.689$^{\mathrm s}$, $\delta_{2000}$=+05$^{\mathrm o}$ 40' 09.25''), located in the field of eclipsing binary DV Psc, was recently reported by Parimucha (2007). He analyzed $V$ and $R_C$ light curves and showed that the system is most probably contact binary with spotted regions on one of the components. Cutri et al. (2003) reported this star as suspected for variability during their 2MASS project. 

\section{Observations and data reduction}

Our photometric observations of the DV Psc field has started on September 4, 2005 and ran to October 26, 2006. The system was detected on CCD images from 5 nights in 2005. Observation field in 2006 was selected in the manner to obtain system during all observations of DV Psc. All these observations were obtained at the Star\'{a} Lesn\'{a} (SL) Observatory of Astronomical Institute of the Slovak Academy of Sciences and at the Kolonica Observatory (KO). The journal of CCD observations is given in Tab.~1.

For observations from SL observatory the 50cm telescope equipped with the SBIG ST-10 MXE CCD camera with standard $BVR_C$ filters was used. Observations from KO were performed by 265/1360 mm Newton type telescope with Meade DSI Pro CCD camera with Sony's ExView HAD Monochrome CCD Image Sensor. No filter was used for these observations.

\begin{table}[t]
\label{tab:journal}
\centering
\caption{The journal of CCD observations of GSC~0008-00901 obtained at Star\'{a} Lesn\'{a} (SL) and Kolonica (KO) observatories. The phase interval is determined according to ephemeris (2). Observations denoted by * were used only for minima times determination.}
\label{tab:1} 
\begin{tabular}{ccccc}
\hline
\hline
Date         & Time (UT)    &     Phase       & Filter & Obs.\\
\hline 
Sep 04, 05  &23:12 -- 03:04 & 0.221 -- 0.598 &$ BVR $  & SL \\ 
Sep 23, 05  &19:45 -- 22:03 & 0.200 -- 0.527 &$ BVR $  & SL \\
Sep 26, 05  &20:39 -- 01:01 & 0.696 -- 0.330 &$ BVR $  & SL \\
Oct 04, 05  &18:36 -- 01:22 & 0.097 -- 0.954 &$ BVR $  & SL \\
Oct 27, 05  &19:21 -- 23:26 & 0.615 -- 0.197 &$ BVR $  & SL \\
Aug 15, 06  &22:35 -- 02:25 & 0.769 -- 0.320 & --      & KO* \\
Aug 17, 06  &22:28 -- 02:26 & 0.661 -- 0.234 & --      & KO* \\
Aug 26, 06  &21:28 -- 02:43 & 0.608 -- 0.363 & --      & KO* \\
Sep 16, 06  &19:02 -- 00:45 & 0.290 -- 0.856 & --      & KO*\\
Oct 17, 06  &18:04 -- 00:54 & 0.751 -- 0.713 &$ BVR $  & SL \\
Oct 18, 06  &19:27 -- 00:57 & 0.405 -- 0.192 &$ BVR $  & SL \\
Oct 26, 06  &18:35 -- 00:12 & 0.914 -- 0.714 &$  VR $  & SL* \\
\hline
\hline
\end{tabular}
\end{table}

The CCD images were reduced in the usual way (bias and dark subtraction, flat-field correction) in software package SPHOTOM developed by the first author. This package was also used for aperture photometry of all detected stars on the CCD frames. This procedure is based on the latest SExtrator code (Bertin and Arnouts, 1996). The brightness of the star was determined with respect to BD~+04$^{\mathrm o}$19 (GSC~00008-00949) with $B$ = 12.04 mag, $V$ = 11.47 mag, $R$~=~11.22 mag, (Monet et al., 2003). The stability of comparison star was tested with respect to GSC~{00008-00743} and BD~+04$^{\mathrm o}$18 and was found to be stable within $\sim$0.003 mag during all observations.

Instrumental magnitudes $b,v$ and $r$ (for SL observations) and $\Delta m$ (for KO observations) were determined. Nightly atmospheric extinction coefficients were computed from the comparison stars measurements and the correction for an atmospheric extinction was applied to these observations. Above-atmosphere instrumental magnitudes $b_0,v_0$ and $r_0$  were determined and consecutively were corrected to the standard $BVR_C$ international system using following equations :

\begin{equation}
\label{transform}
\centering
\begin{array}{rcl}
V        &   =   &  v_0 + k_{v}(B-V)\\
B - V    &   =   &            k_{bv} (b_0 - v_0)  \\
V - R_C    &   =   &            k_{vr} (v_0 - r_0), \\
\end{array}
\end{equation}

\begin{table}
\centering
\label{minima:table}
\caption{Weighted average $BVR$ times of the primary (I) and secondary (II) minima
of GSC 00008-00901. The standard errors of the minima are given in parentheses.}
\begin{tabular}{ccc}
\hline
\hline
HJD & Type & Filter\\
2400000+ & &\\
\hline
53618.5935(2) &II&$BVR$\\
53640.4491(1) &I &$BVR$\\
53648.4064(2) &II&$BVR$\\
53671.4278(2) &I &$BVR$\\
53963.5096(1) &I &none\\
53965.5402(1) &I &none\\
53974.5140(2) &I &none\\
53995.3626(5) &I &none\\
54026.3318(2) &I &$BVR$\\
54026.4709(2) &II&$BVR$\\
54027.3427(2) &II&$BVR$\\
54027.4921(1) &I &$BVR$\\
54035.3076(2) &I &$VR $\\
54035.4492(1) &II&$VR $\\
\hline
\hline
\end{tabular}
\end{table}

\noindent where $k_v$=0, $k_{bv}$=0.834 and $k_{vr}$=1.467. Average transformation coefficients are routinely determined one or twice per month from measurements of the Landolt standarts (Landolt, 1992). KO observations were not reduced to international system and were used for minima times determination only (see Section~3). 

GSC 00008-00901 is relatively faint (with $V_{max} \sim 13.9$ mag) star and exposure times of our observations in different passbands were selected in focus to DV~PSc. It strongly affected a quality of the obtained light curves of this system with our instruments. Because of poor quality of photometric data, we had to remove from light curves analysis SL observations from October 26, 2006. This observation was suitable only for minima times determination.

The average accuracy of the observations in $B$ and $V$ passband is $\sim$0.005 mag and in $R_C$ passband $\sim$0.008 mag. The accuracy of the KO observations is about 0.02 mag.

\begin{figure}
\centering
\includegraphics[width=0.5\textwidth]{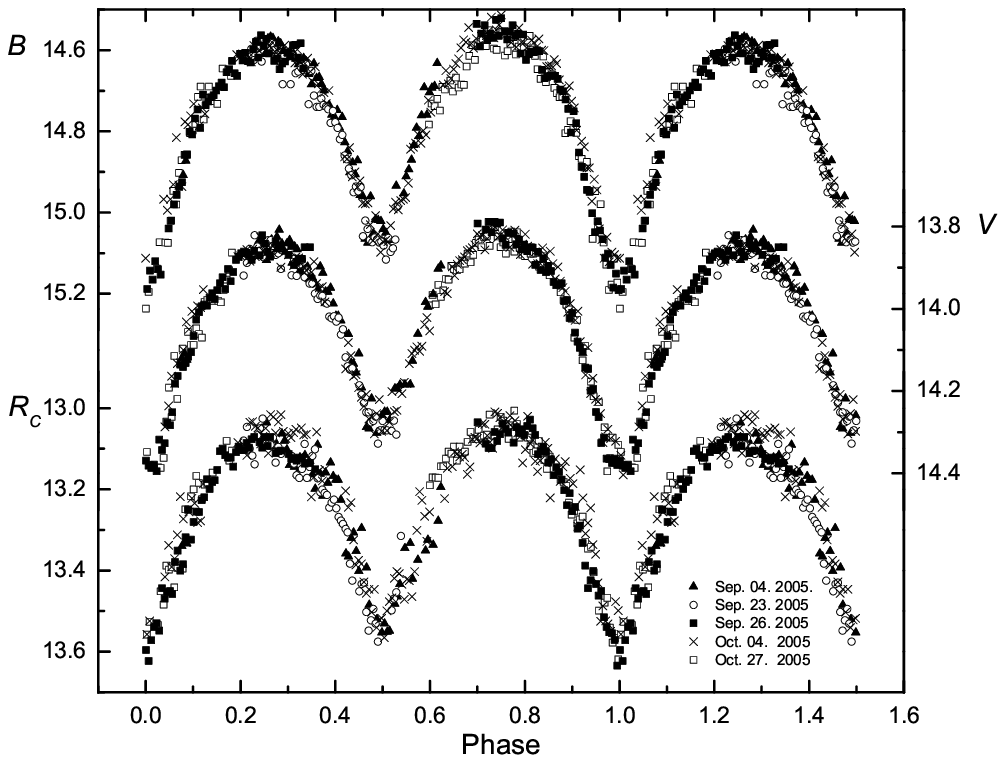}
\includegraphics[width=0.5\textwidth]{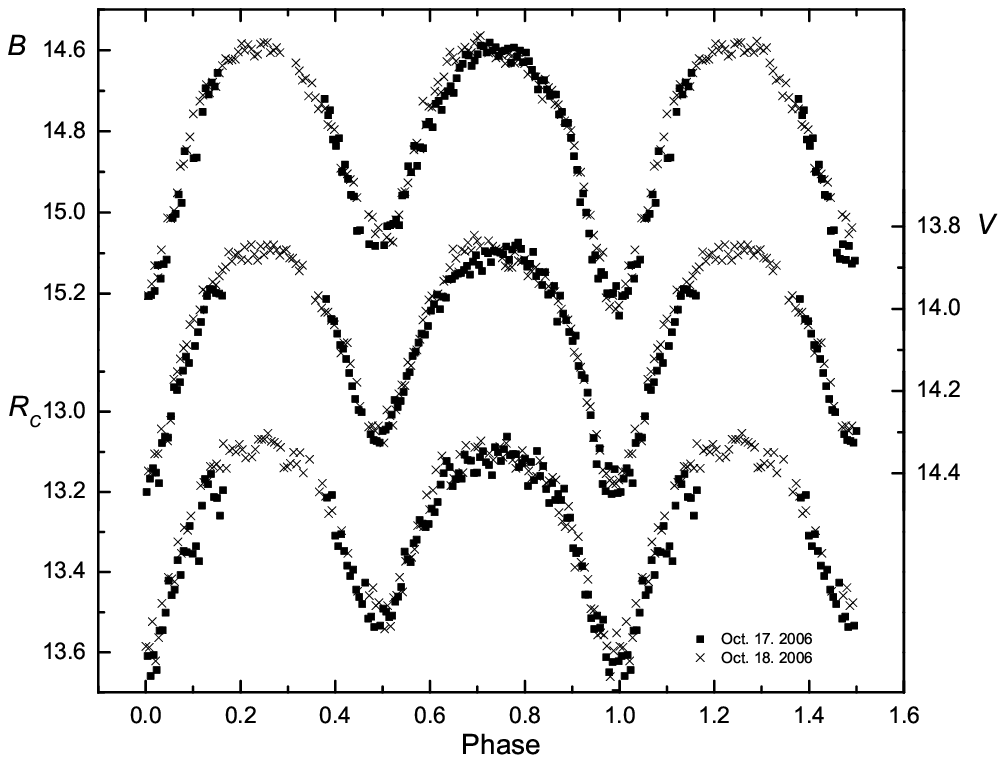}
\caption{The $BVR_C$ light curves of GSC 00008-00901 obtained in 2005 (top) and 2006 (bottom).}
\end{figure}

\section{Minima times and period determination}

Our observations enabled us to determine totally 14 minima times of GSC~00008-00901. They are listed in Tab.~2. The times of minima were determined separately for all filters (SL observations) using Kwee and van Woerden (1956) method and then weighted average minima from all filters were determined. For computation of minima times we have used only data in a phase interval $\pm$0.05 around the minimum. This approach minimize the influence of the minima asymmetries (van't Veer, 1973).

Using period analysis of the light curves during discovery process, the period $P$ = 0.289437(29) days was found. A brief inspection of the light curve obtained on October 4, 2005 showed that observed minimum is secondary. It enables us to estimate minima types as listed in Tab.~2.

Listed minima times were used to improve period and provide linear ephemeris for primary minimum:

\begin{equation}
\centering
\label{ephem:1}
\begin{array}{rrl}
 \mathrm {Min~I} = 2453648.26421 & +~0.28948 & \times E,\\
                \pm 23  &     \pm 11   &\\
\end{array}
\end{equation}
which was used to phase all our data. 

The $BVR_C$ light curves of GSC~00008-00901 obtained in 2005 and 2006 seasons, phased with ephemeris (\ref{ephem:1}) are depicted in Fig.~1.

\section{Light curves analysis}

A more detailed look at light curves of GSC~00008-00901 obtained in 2005 and 2006 seasons shows that they are quite different. Observations from 2006 show symmetric light curves in all passbands
with brightness in the both maxima at the same level. On the other hand, data from 2005 indicate that the brightness of the system in maximum in phase 0.75 (MaxII) is brighter than maximum in phase 0.25 (MaxI).To quantify this effect, we performed quadratic polynomial fit of the observed data around the MaxI and MaxII and we found that MaxI-MaxII is 0.033 mag for $B$ filter, 0.028 mag for $V$ and 0.02 mag for $R_C$ filter, respectively. This indicates a presence of a spot(s) on the surface of the stars. So, we decided to solve light curves from this seasons separately.

The analysis of the light curves  was performed by software package PHOEBE (Pr\v{s}a and Zwitter, 2005). It is a modeling package for eclipsing binary stars, which is based on widely used WD program (Wilson and Devinney, 1971; Wilson, 1979; 1990).

\begin{figure}
\centering
\includegraphics[width=0.5\textwidth]{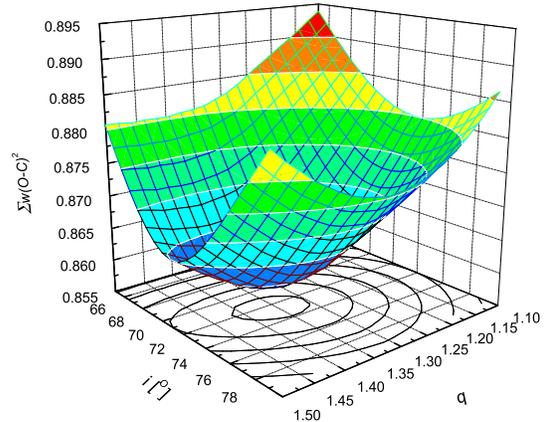}
\caption{The variations of the normalized weighted sum of squares of residuals $\sum w (O-C)^2$ for all light curves from 2006 for different values of mass ratio $q$ and orbital inclination $i$.}
\end{figure}

In the process of light curve analysis of the binary we have adopted and fixed following parameters: the mean temperature of the primary component, $T_1 = 5350 K$, based on the $B-V$=0.72$\pm$0.09 mag color index, which corresponds to G8 spectral type (Cox, 2000), coefficients of gravity darkening  $g_1 = g_2 = 0.32$ (Lucy, 1967) and the bolometric albedo coefficients of $A_1 = A_2 = 0.5$ (Rucinski, 1969), which are appropriate values for the convective envelopes. A Kurucz (1993) model of stellar atmospheres was applied to the stars assuming solar composition. The second order bolometric and monochromatic limb darkening coefficients for logarithmic law, were interpolated from van Hamme (1993) tables.

The weights of individual data points were established as $1/\sigma^2$, where $\sigma$ is standard error of point derived during photometric measurement. In our solutions, we used all data points, we have not calculated normal points as usual. Using this approach we do not loose information in our data. 

We fitted all three light curves simultaneously. Based on appearance of the light curves, we set PHOEBE solution to overcontact binary, which is not in thermal contact. Initial values of free parameters were determined manually to give a visually good fit with our data. With these values, a correct solution was quickly found. We have run differential correction fit as long as output corrections were smaller than the errors of the fitted parameters. 

\begin{table}
\centering
\caption{Photometric elements of GSC 00008-00901 and their standard errors (in
parenthesis) for 2006 light curves: $i$ - orbital inclination, $q$ - photometric mass ratio, $\Omega$ - surface potential, $f$ - fill-out factor, $r_1, r_2$ -volume mean fractional radii, $T_1$ - fixed temperature of the primary component, $T_2$ - temperature of secondary component, $\sum w (O-C)^2$ - normalized weighted sum of squares of residuals for all light curves}
\begin{tabular}{lc}
\hline
\hline
Parameter &   \\
\hline
$i[^o]$             & 72.35(15)\\
$q[^o]$             & 1.332(11)\\
$\Omega$            & 4.194(15)\\
$f$                 & 0.114(26)\\
$r_1$               & 0.359(5)\\
$r_2$               & 0.417(9)\\
$T_1[K]$            & 5350a\\
$T_2[K]$            & 4995(8)\\
$L_1/(L_1+L_2)_B$   &  0.557(30)\\
$L_1/(L_1+L_2)_V$   &  0.530(28)\\
$L_1/(L_1+L_2)_{R_C}$ & 0.513(25)\\
\hline 
$\sum w (O-C)^2$   &  0.859\\
\hline 
\hline
\end{tabular}
\end{table}

\begin{table}
\centering
\caption{Temperature of the secondary component $T_2$ and spot parameters and their standard errors (in
parenthesis) for different configurations: CP - cool spot on primary, CS - cool spot on secondary, HP - hot spot on primary, HS - hot spot on secondary, $lon$ - longitude of the spot, $lat$ - latitude of the spot, $rad$ - radius of the spot, $k$ - temperature factor (fixed), $\sum w (O-C)^2$ - normalized weighted sum of squares of residuals for all light curves }
\begin{tabular}{lllll}
\hline
\hline
Parameter         &   CP      &     CS     & HP      &  HS\\ 
\hline
$T_2[K]$          &  5158(8) &   5171(5)   & 5162(7) & 5169(9) \\
$lon[^o]$         &  282(5)   &   112(4)   & 105(6)  & 285(5)  \\
$lat[^o]$         &  71(13)   &   73(14)   & 70(14)  & 71(11)  \\
$rad[^o]$         &  26(2)    &   30(3)    & 22(3)   & 21(2)   \\
$k$               &   0.95a   &   0.95a    &  1.05a  &  1.05a \\
\hline
$\sum w (O-C)^2$  &   2.235   &   2.214    & 2.236   & 2.229  \\
\hline
\hline
\end{tabular}
\end{table}

In the first step, we solved light curves from 2006, when we assumed that there are no spots on the surface of the stars.
According to Binnendijk (1957) dividing of W UMa systems into A and W types, we concluded that this system is a W-type.
because of G8 spectral type and short orbital period. It means, that primary minimum is caused by occultation of the less massive and hotter primary star by more massive but cooler secondary component. Beacause of W-type definition, the mass ratio $q>1$.

To find correct values of mass ratio $q$ and orbital inclination $i$, we computed normalized weighted sum of squares of residuals $\sum w (O-C)^2$ for fixed values of $q$ in interval 1.0--1.8 and $i$ in interval 60--90$^o$ and looked for a global minimum. A detailed view of the variations of the sum of squares around the global minimum is depicted at Fig~2. We obtained photometric parameters, which are listed in Tab.~3. The best fits with corresponding O-C for all passbands are depicted in Fig.~3 (bottom).

These parameters were used in the next solution of 2005 light curves. As was mentioned before, the light curves from this season are affected by maculation. The process of the searching for the spot's parameters  is complicated by the fact that observed O'Connell effect could be caused by different combinations of the spot's location (on primary or secondary component) and spot's temperature (hot or cool spot). To quantify this effect, we tried to fit light curves from 2005 for 4 configurations of the spot location and temperature as listed in Tab.~4. In these solutions we have fixed temperature factor (ratio between temperature of the spot and surrounding atmosphere) of the spot as well as parameters from 2006 solution except temperature of secondary component. As could be seen from values of $\sum w (O-C)^2$ in Tab.~4, to select correct solution is practically impossible. In Fig.~3 (top) we show light curve solution in the case when a cool spot is located on the secondary component. This solution has a minimal weighted sum of squares of residuals for all light curves. The corresponding 3D model of the system with a cool spot on secondary component is shown in Fig.~4.

\begin{figure}
\centering
\includegraphics[width=0.5\textwidth]{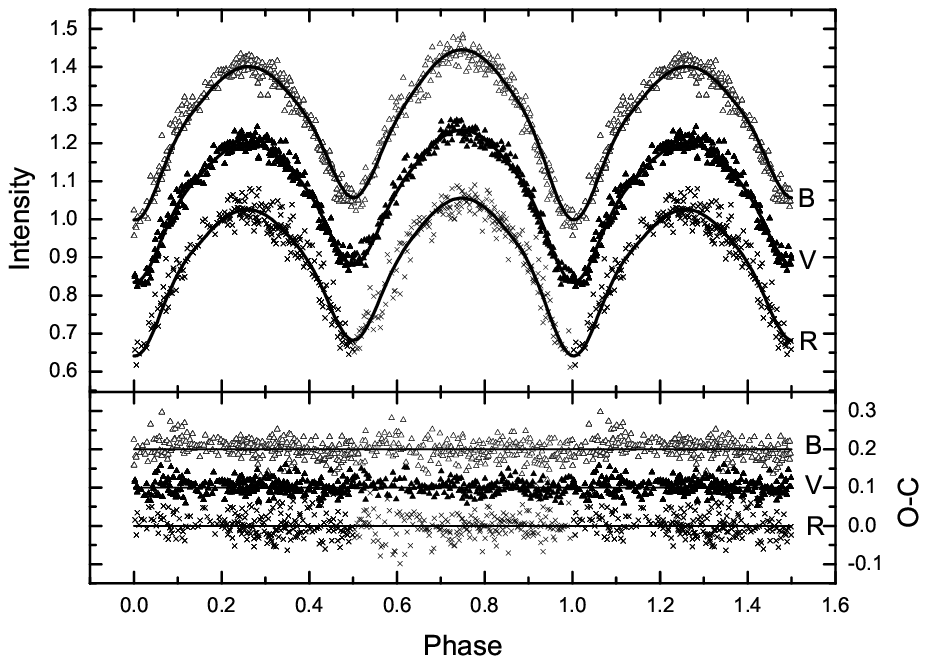}
\includegraphics[width=0.5\textwidth]{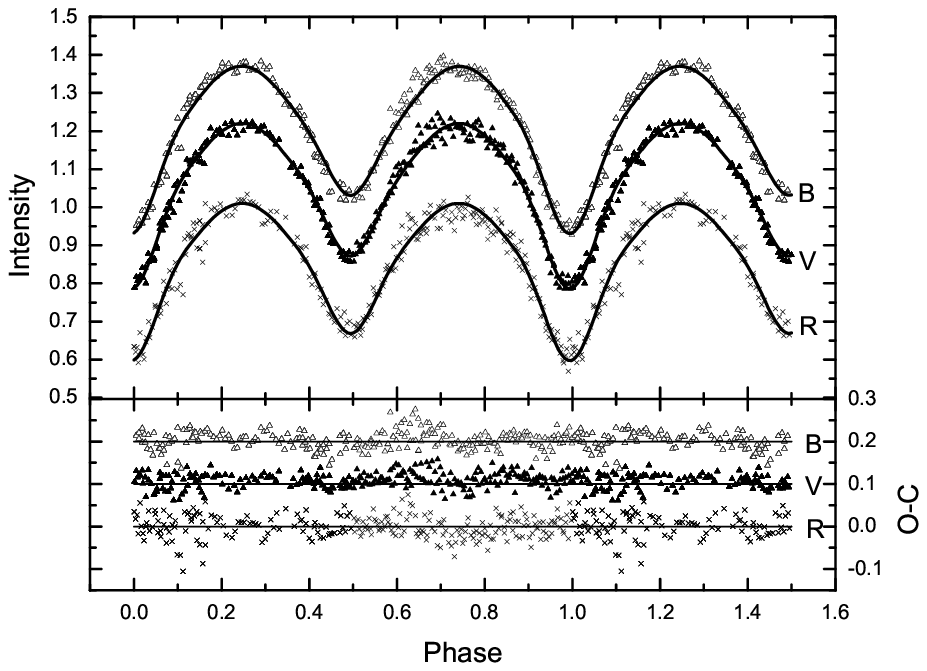}
\caption{The best fits of $BVR_C$ light curves of GSC 00008-00901 for 2005 data (top) and 2006 data (bottom). $B$ and $V$ data are shifted for a clarity.}
\end{figure}

\section{Discussion and Conclusion}

Our light curve analysis of $BVR_C$ photometric observations of recently discovered eclipsing binary GSC 00008-00901, which is located in the field of another eclipsing binary DV~Psc showed, that this system is an overcontact binary of W~UMa type. Considering of its spectral type and orbital period, we concluded that the primary minimum is occultation (primary component is eclipsed by secondary one). So, this system belongs to an W-type of W UMa binaries. The temperature difference between components is about 350K.

In spite of problems with quality of the observations we showed that light curves from the both seasons 2005 and 2006 are different. The light curves from 2005 show different brightness in the both maxima, while 2006 light curves are symmetric.
Asymmetry of the light curves from 2005 could be explained by the spot on one of the components, however, some conclusions on spot's temperature and/or its location (on primary or secondary component) are uncertain.

\begin{figure}
\centering
\includegraphics[width=0.49\textwidth]{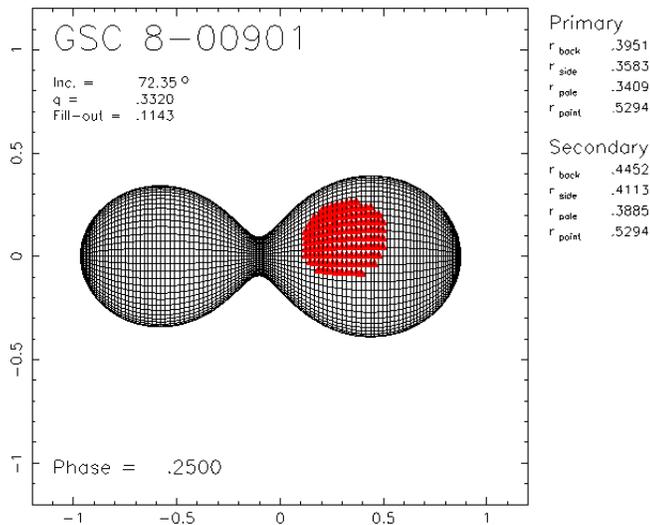}
\caption{3D model of GSC 00008-00901 with a cool spot on the secondary component}
\end{figure}

\begin{acknowledgements}
This paper was supported by these grants: VEGA grants of the Slovak Academy of Sciences No. 7010 and 7011, grants of the \v{S}af\'arik University VVGS 10/2006 and VVGS 9/07-08, APVV grant LPP-0049-06, Bilateral APVV grant SK-UK-01006,  INTERREG IIIA SR-\v{C}R 143-13-36 grant and APVT grant 20-014402. M, Vanko's research is supported by a Marie Curie "Transfer of Knowledge" Fellowship within the 6th European Community Framework Programme.
\end{acknowledgements}

\end{document}